\documentclass[%
 twocolumn,
superscriptaddress,
 amsmath,amssymb,
 aps,
]{revtex4}

\usepackage{graphicx,color}
\usepackage{dcolumn}
\usepackage{bm}

\usepackage{mwe} 
\usepackage[caption=false]{subfig}

\begin{document}

\preprint{APS/123-QED}

\title{Artificial polariton molecules}

\author{Alexander Johnston}
    \affiliation{Department of Applied Mathematics and Theoretical Physics, University of Cambridge, Wilberforce Road, Cambridge, CB30WA, United Kingdom}
\author{Kirill P. Kalinin}
    \affiliation{Department of Applied Mathematics and Theoretical Physics, University of Cambridge, Wilberforce Road, Cambridge, CB30WA, United Kingdom}
\author{Natalia G. Berloff} 
\email[correspondence address: ]{N.G.Berloff@damtp.cam.ac.uk}
    \affiliation{Department of Applied Mathematics and Theoretical Physics, University of Cambridge, Wilberforce Road, Cambridge, CB30WA, United Kingdom}
     \affiliation{Skolkovo Institute of Science and Technology, Bolshoy Boulevard 30, bld. 1
Moscow, 121205, Russia }

\date{\today}

\begin{abstract} We show that geometrically coupled polariton condensates fabricated in semiconductor devices are versatile systems capable of simulating molecules with given characteristics. In particular, we consider oscillatory and stationary symmetric and asymmetric states in polariton dimers, trimers, and tetrads and their luminosity in real and Fourier space. The spectral weights of oscillatory states are associated with discrete spectral lines. Their number and separation can be controlled by changing the number and geometry of condensates, reflected by the coupling strengths. We also show that asymmetric stationary states combine discrete and continuous degrees of freedom  in one system. The continuous degree of freedom is represented by the phase while the discrete degree of freedom is given by density asymmetry. Our work paves the way to engineer  controllable artificial molecules with a range of properties manufactured on demand.

\end{abstract}

\maketitle

The complex behaviour of many-body solid-state and photonic systems provides a rich platform for exploring disparate physical behaviours while opening up the ability to predict and create new classes of materials with designed properties. The notion of an  ``artificial atom" suggests that some nanostructures have  size-dependent physical properties. For instance,  colloidal nanoparticles of inorganic solids or quantum dots with their controllable sizes and characterisation enable the periodic table to acquire a third dimension \cite{choi2010artificial}. This analogy can be developed further by showing that artificial atoms can be grouped and coupled together to create  ``artificial molecules". The physical properties of such molecules, including their electronic, optical, and mechanic properties are drastically different from those of the component atoms due to interaction couplings. How can we build a system that allows us to control couplings between elementary atoms to simulate the behaviour of artificial molecules?   Well-defined groupings of  coupled constituents such as  neutral atoms, ions, superconducting circuits, quantum dots, nanocrystals, nuclear spins, photons, polaritons, or polariton condensates a may show properties altered from those of the individual components and form artificial molecules.  On the one hand, such artificial molecules  may be capable of simulating a wide range of elaborate Hamiltonians \cite{georgescu2014quantum}. Recently, optical and polaritonic lattices have been proposed as models for an analogue simulator capable of finding the lowest energy states of spin Hamiltonians such as Ising and XY, as well as higher order $k$-local  Hamiltonians \cite{ amo2016exciton,rechcinska2019engineering,parto2020realizing, pal2019rapid,  BerloffNatMat2017, suchomel2018platform, stroev2019discrete,kalinin2018networks}. On the other hand, an optical or laser element, nanocrystal, or polariton or photon condensate all have more degrees of freedom than just ``spin", which is in these systems associated with the phase of the order parameter that describes such artificial atoms. 
In this Letter, we show that coupled photonic or polaritonic systems form the clusters while the controllable interactions between constituents play the role of molecular bonds giving rise to new energy states, optical properties, and vibrations that can be spectroscopically observed.

Polaritons  are  bosonic quasiparticles formed by blending light (photons) and matter (excitons) in semiconductor microcavities \cite{carusotto2013quantum,keeling2011exciton}.  They  form a Bose-Einstein condensate (BEC) at higher temperatures than atomic systems \cite{kasprzak2006bose}. Lattices of polariton condensates are able to interact through the outflow of polaritons from each condensate center (CC) \cite{ohadi2016nontrivial}. The interaction between condensates, facilitated by the coupling, results in the emergence of synchronisation and phase-locking \cite{baas2008synchronized,wouters2008synchronized}. Networks of polariton condensates were shown to relate to  Kuramoto, Sakaguchi-Kuramoto, Stuart-Landau, and Lang-Kobayashi oscillators and beyond \cite{kalinin2019polaritonic}. At the same time, polariton condensates differ from optical (laser) systems by the presence of self-interactions. As we establish in this Letter, such nonlinearities lead to novel states that can be utilised to combine discrete and continuous degrees of freedom and to change  the form of  artificial molecule bonds and deformations.

The ability to design and synthesise polariton artificial molecules is interesting for predicting some new behaviours. As the couplings between the individual condensates change, the resonance splits into lower and higher frequency modes. This is analogous to plasmon coupling in trimers and quadramers of metal nanoparticles \cite{brandl2005plasmon} and hollow metal nanospheres \cite{prodan2003hybridization}. Each of the constituents in these systems is an oscillator  that has a well-defined amplitude and density. Photonic molecules have been shown to display novel quantum optical behaviours, making them useful tools in the construction of photonic devices \cite{liao2020photonic}.  They have  also been observed in polariton micropillar structures \cite{galbiati2012polariton,ferrier2010polariton,abbarchi2013macroscopic}, with couplings   greatly influencing the nature of the condensation. 

The coupling of two polaritonic/photonic (nonequilibrium) CCs has been studied  in the framework of the weak lasing regime \cite{nelsen2009lasing,tosi2012sculpting,aleiner2012radiative,st2017lasing,kassenberg2020controllable}, demonstrating the existence of  steady, oscillatory and chaotic states.
Two CCs can emit a  number of equidistant frequencies simultaneously forming a frequency comb \cite{rayanov2015frequency,ruiz2020autonomous,kim2020emergence}, a phenomenon comparable to those observed in mode-locked lasers \cite{udem2002optical,cundiff2003colloquium} and optical microresonators \cite{del2007optical,kippenberg2011microresonator,kassenberg2020controllable}. 

 In this Letter we show that polariton condensates arranged in simple geometrical configurations of several elements are flexible emulators of structural, spectral, and physical properties associated with complex molecules; oscillatory states can be used to control the spectral gaps of the emission whereas asymmetric stationary states can be used to combine discrete and continuous degrees of freedom in one molecule.
 
 Our starting point is a general gain-saturation model that describes a system of coupled oscillators such as lasers or nonequilibrium condensates: 

\begin{equation}
\label{wave_function_rate_equation_simplified}
\dot{\psi_{i}} = \psi_{i}\biggl(\frac{P}{1 + |{\psi_{i}}|^2} - 1\biggr)- i s |\psi_i|^2 \psi_i + (1 - ig)\sum_{j \neq i} J_{ij}\psi_{j},
\end{equation}
where $\psi_i(t)=\sqrt{\rho_i}\exp[i \theta_i]$ is the complex amplitude of the $i-$th CC, $\rho_i$ and $\theta_i$ are occupation and phase, respectively, $s$ is the strength of the nonlinearity (self-interactions) within each CC, $P$ is the pumping strength, $g$ is the strength of the detuning, and $J_{ij}$ is the coupling strength between the $i-$th and $j-$th CCs.  These equations were derived from the full mean-field Maxwell-Block equations for laser cavities \cite{dunlop1996generalized} or using the tight-binding approximation from the mean-field complex Ginzburg-Landau equation in the fast reservoir regime \cite{kalinin2019polaritonic}. For lasers $s=0$, whereas for nonequilibrium condensates $s$ can be positive or negative depending on the governing system parameters. The coupling strength between each pair of CCs, $J_{ij}$, depends on the pumping strength and on other system parameters such as the distance between the condensates or the shape of the pumping beam \cite{ohadi2016nontrivial}.

For two CCs and $s=0$ we can obtain the fully analytical steady state solutions of Eq.~(\ref{wave_function_rate_equation_simplified}) written in terms of occupations and phases:
\begin{eqnarray}
\label{rho_dot_simplified}
    0 &=& \Big(\frac{P}{1 + \rho_{i}} - 1\Big)\cos \delta + J\sqrt{\frac{\rho_{j}}{\rho_{i}}}\cos({\theta_{ij}} +\delta), \\
\label{theta_dot_simplified}
 \mu &=&-s \rho_i -J\sqrt{\frac{\rho_{j}}{\rho_{i}}}\sin({\theta_{ij}} + \delta)/\cos\delta,
\end{eqnarray}
where we denoted $\theta_{ij} \equiv \theta_{i} - \theta_{j}$, $\theta=\theta_{12}$, $\tan{\delta} = g$, $J_{12}=J$  and introduced the chemical potential $\mu$ that characterises the global frequency of the system. We introduce the density discrepancy  $F\ge 1$ by writing $\rho_2=F^2 \rho_1$. It follows from Eqs.~(\ref{rho_dot_simplified}-\ref{theta_dot_simplified}) that  $F=1$ if and only if   $\theta=0$ or $\pi$. 
For asymmetric solutions ($F>1$)  and $s=0$ we have
\begin{eqnarray}
\label{rho_ratio_simplified_equation}
    \frac{\rho_{2}}{\rho_{1}} &\equiv &F^2(\theta) =  \frac{\sin{(\delta - \theta)}}{\sin{(\delta + \theta)}},\\
    \rho_1&=&\frac{J \cos \theta}{J \sin(\theta - \delta)\sin \delta - F \cos^2 \delta}. \label{density_equation}
    \end{eqnarray}
 We substitute Eqs.~(\ref{rho_ratio_simplified_equation}) and (\ref{density_equation}) into Eq.~(\ref{rho_dot_simplified}) and solve it for $P$ in terms of $\theta$ for fixed $J$ and $0<\delta<\pi/2$. This gives a family of asymmetric solutions, such that each $\theta$ determines the pumping $P$ that leads to the solution with that $\theta$ if $P,\rho_1>0$ and $F>1$. These conditions are satisfied for  $-\delta < \theta <0$ if $J<0$ and for $\pi-\delta < \theta < \pi$ if $J>0$. 
 
\begin{figure}[h]
 \includegraphics[width=\columnwidth]{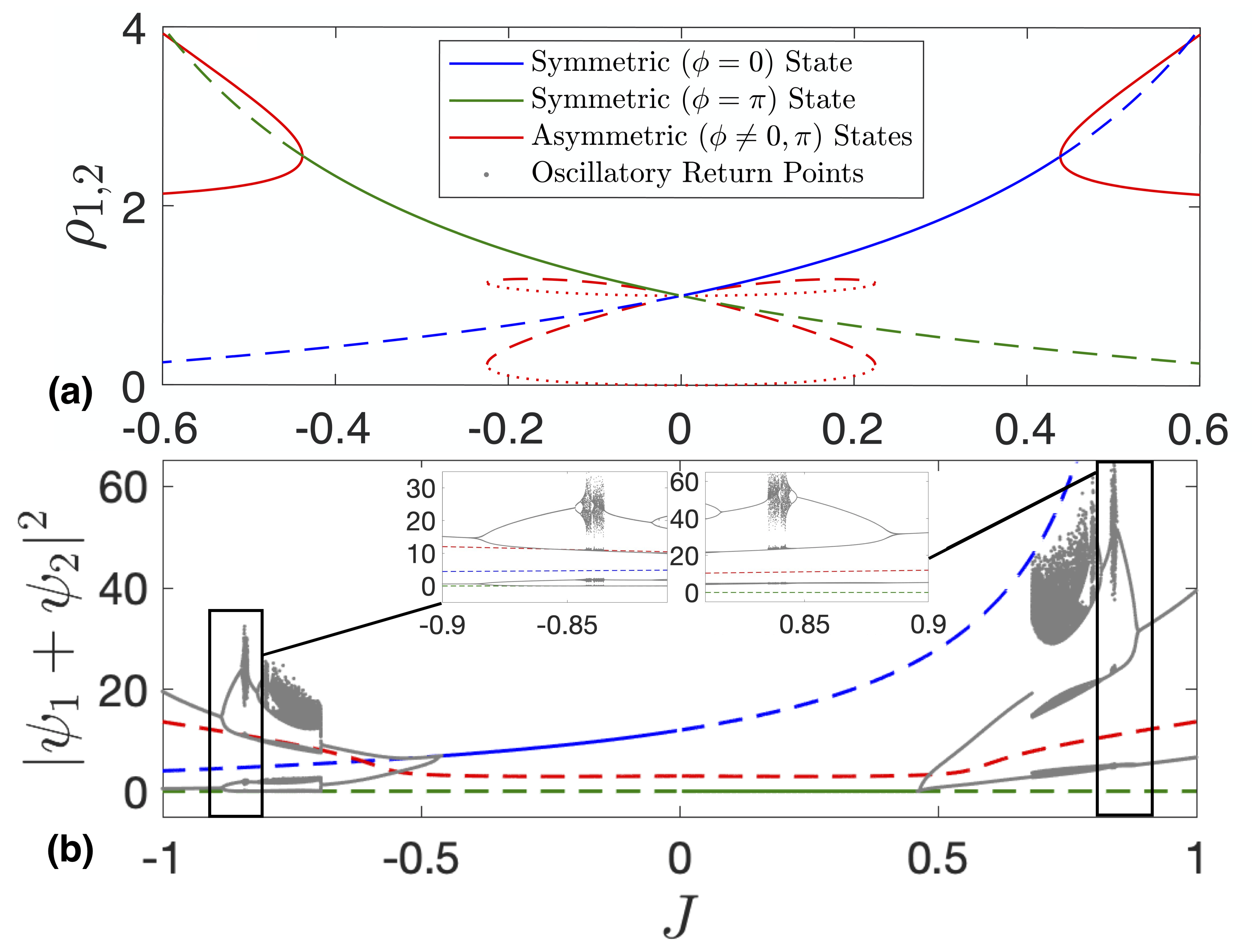}
   \caption{Bifurcation diagrams of polariton dimers for $P=2$ (a) and $P=4$ (b) obtained by integrating Eq.~(\ref{wave_function_rate_equation_simplified}) for $g = 1$ and $s = 0.5$.  In (a) there are no oscillatory states and we plot the individual condensate densities against $J$. In (b) there are oscillatory states and we plot the total intensity $|\psi_{1} + \psi_{2}|^2$ against $J$.  In oscillatory regions the return points (local maxima and minima) of the oscillations are plotted.Two insets illustrate the nature of the period-doubling bifurcations as $|J|$ is decreased. Stable (unstable) states are indicated by solid (dashed and dotted) lines.   }  
  \label{phasediagramDimer}
\end{figure}

  Asymmetric dimers are unstable if $s=0$ and 
  it is the existence of nonlinear self-interactions, encapsulated by the parameter $s$, that are responsible for the emergence of stable asymmetric states, as we show in the Supplementary Information using linear stability analysis. As one changes the parameters of the system, the polariton dimer acquires one of the expected states of the dynamical system: symmetric stationary (with $0$ or $\pi$ phase difference), asymmetric stationary (with $\rho_1\ne \rho_2, $ $\theta\ne 0, \pi$), oscillatory, or chaotic.
Figure~\ref{phasediagramDimer} represents the range of such solutions for a polariton dimer (at two different values of the pumping strength) as the coupling between CCs is varied.  For smaller pumping strengths ($P=2,$  Fig.~\ref{phasediagramDimer}(a)) we see stable symmetric bonding (when $J>0$) or antibonding (when $J<0$) states for lower values of $|J|$ and the emergence of stable asymmetric states at higher values of $|J|$. For larger pumping strengths ($P=4$, Fig.~\ref{phasediagramDimer}(b)) the stable bonding (antibonding) states occur when $J<0$ ($J>0$). These states lose stability at higher $|J|$, transitioning the system into an oscillatory regime.

\begin{figure}[h]
     \includegraphics[width=\columnwidth]{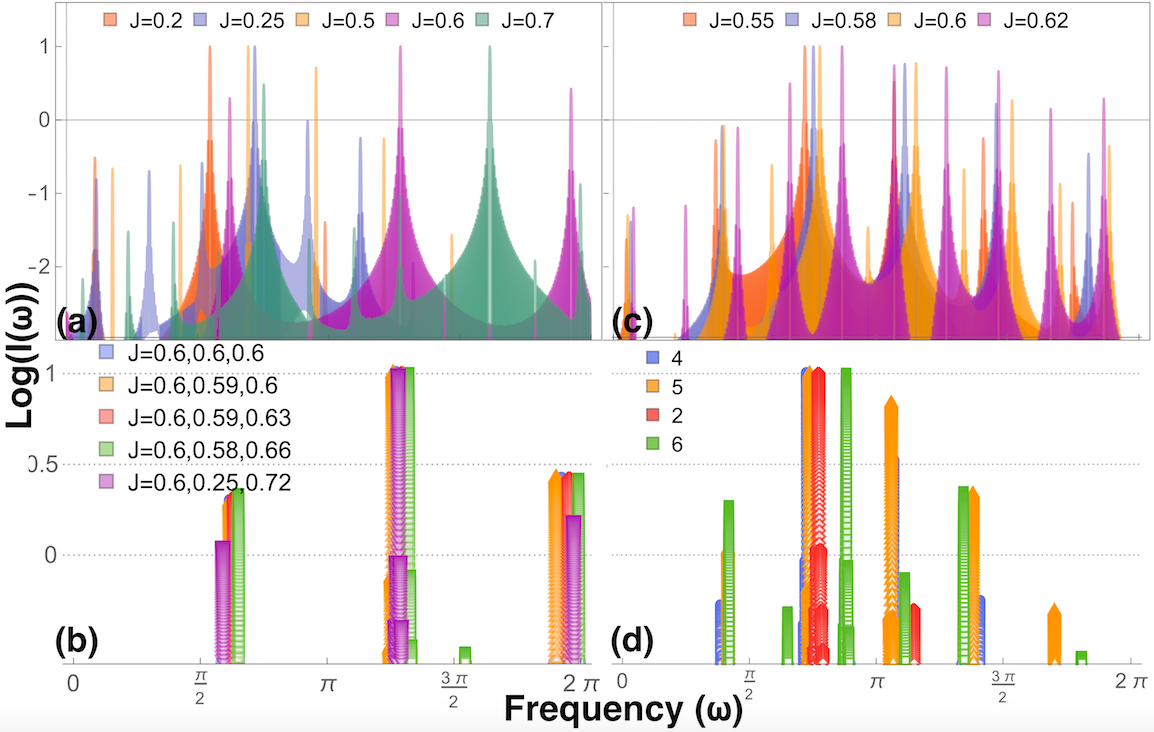}
     \caption{Normalised logarithms of the spectral weight $I(\omega)$ as a function of $\omega$ for trimers (a,b) and tetrads (c,d). (a) All coupling strengths in the triad are the same as given in the legend; (b) all coupling strengths in the trimer are varied as shown in the legend; (c) coupling strengths are $J_{12}=J_{23}=J_{34}=J_{41}=J$ as given in the legend, whereas $J_{24}=J_{13}=0$;  (d) the coupling strengths for blue, yellow, red, and green plots are $J_{i-1,i+1}=0$ and  $J_{i,i+1}=0.55$, $J_{i,i+1}=0.55, 0.57, 0.55, 0.55$, $J_{i,i+1}=0.55, 0.58, 0.61, 0.63, $  and $J_{i,i+1}=0.55, 0.58, 0.7, 0.3,$  respectively. The number of significant spectral lines changes depending on the couplings. In all plots $g=1,s=0.5$ in Eq.~(\ref{wave_function_rate_equation_simplified}); (a,b) $P=3$; (c,d) $P=1$.}
    \label{spectra}
\end{figure}

By increasing the number of condensates with a concomitant increase in the variability of couplings between them, we can achieve the same types of behaviours (stationary, oscillatory, or chaotic) with extra flexibility in the control of the structural properties of the states. With polariton trimers and tetrads we can tune the spectral gaps  and the number of spectral lines on demand, as Fig.~\ref{spectra} illustrates. 
In experiments, the state of a polariton system is probed by analysing the momentum- and energy-resolved photoluminescence spectrum that can be directly measured in the far field. The spectral weight is given by the modulus squared of the Fourier transform of the wave function, 
\begin{equation}
I(\omega) = \biggl | \int e^{-i \omega t} \psi(t) dt\biggr |^2 = \biggl | \int e^{-i \omega t} \biggl[\sum_{i=1}^N \sqrt{\rho_i} e^{i \theta_i}\biggr] dt\biggr |^2.
\label{sw}
\end{equation}
Figure~\ref{spectra} depicts the peaks of the spectral weight for oscillatory trimers (Fig.~\ref{spectra}(a,b)) and tetrads (Fig.~\ref{spectra}(c,d)) if the condensates have the same (Fig.~\ref{spectra}(a,c))  or different (Fig.~\ref{spectra}(b,d)) coupling strengths. Figure~\ref{spectra}(b) shows that by varying the coupling strengths between constituents in a trimer  one can change the spacing (spectral gap) between the spectral lines with the trimer representing a three-level system - a  basic element of universal quantum information processing \cite{aharon2013general,boulier2014polariton,li2017probing}. In addition to changing the distance between the levels, it is possible to change the number of lines by going to a polariton tetrad with an increased number of degrees of freedom, as Fig.~\ref{spectra}(d) illustrates.

\begin{figure}[h!]
     \includegraphics[width=\columnwidth]{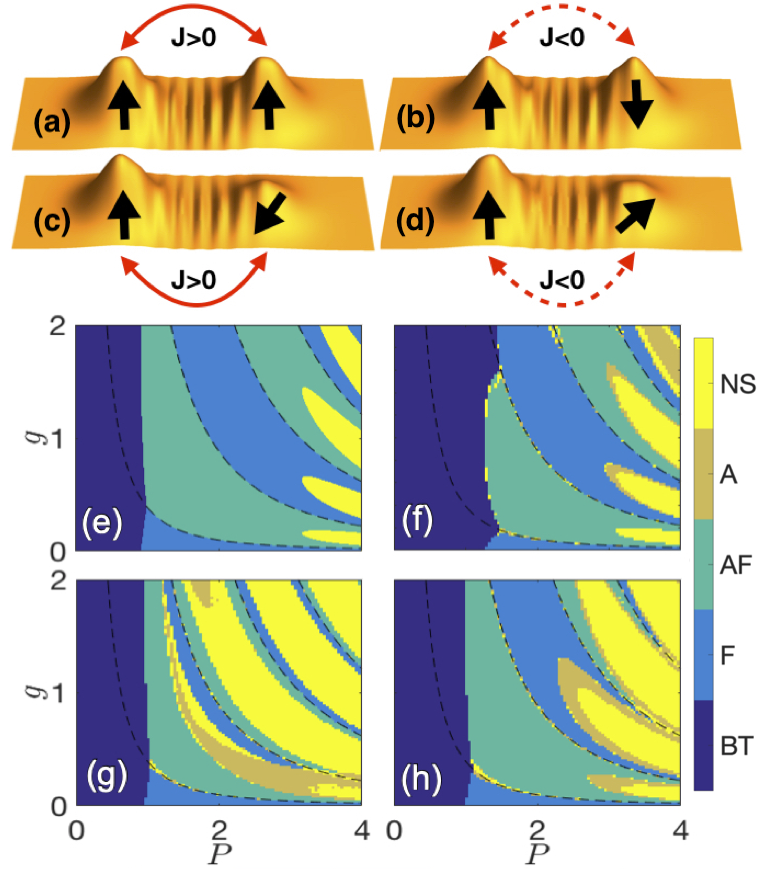}
     \caption{(a)-(d): Schematics of bonding symmetric (a), antibonding symmetric (b), and asymmetric (c,d) states for a dimer. The height (arrows) of the 3-D plot represents the condensate density (phase). (e)-(f): Bifurcation diagrams showing regions  below threshold (BT), ``ferromagnetic" (bonding) (F), ``antiferromagnetic" (antibonding) (AF), asymmetric (A), and non-stationary (NS) states. Lines $J=0$ are shown by the dashed black lines. The diagram in (e) was calculated analytically from  Eq.~(\ref{wave_function_rate_equation_simplified}) with $s=0$.  Asymmetric  states were found for $s=1$, as shown in  (g). (f) and (h) show diagrams obtained numerically from Eqs.~(\ref{wave_function_rate_equation}-\ref{reservoir_dynamics_equation})  for $b_{0} = 0.1$ (f) and $b_{0} = 0.5$ (h), respectively. Also, $\beta = 2.5$, $J_{0} = 0.1$ and $b_1=1$.}  
    \label{dyad_figure}
\end{figure}



Equation~(\ref{wave_function_rate_equation_simplified}) is a simplified model so next we consider how  more realistic rate equations for a particular system would change the stability properties of the artificial molecules. In particular, for a network of $N$ optically excited and freely expanding condensates in the tight-binding approximation \cite{kalinin2019polaritonic} we have a system of $N$ equations on condensate wavefunctions,
\begin{equation}
\label{wave_function_rate_equation}
\dot{\psi_{i}} = -i |{\psi_{i}}|^2\psi_{i} - \psi_{i} + (1 - ig)[R_{i}\psi_{i} + \sum_{j \neq i} J_{ij}\psi_{j}],
\end{equation}
coupled to the rate equations on the hot exciton reservoir densities $R_i$,
\begin{equation}
\label{reservoir_dynamics_equation}
\dot{R_{i}} = b_{0}(P - R_{i} - \xi  R_{i}|{\psi_{i}}|^2).
\end{equation}
Here, in addition to other terms present in Eq.~(\ref{wave_function_rate_equation_simplified}), we have  $\xi = b_{1}/b_{0}$, where $b_0$ and $b_1$ characterise the relaxation rate of the reservoir and the strength of the nonlinearity saturation, respectively. Close to the condensation threshold and in fast reservoir relaxation limit $b_0\gg 1$, Eqs.~(\ref{wave_function_rate_equation}-\ref{reservoir_dynamics_equation}) and Eq.~(\ref{wave_function_rate_equation_simplified})   behave similarly as can be seen by rescaling $\psi_i\rightarrow \tilde{\psi_i}/\sqrt{\xi}$, assigning $s=gP-1$, and taking the Taylor expansion in the expression for $R_i=P/(1+|\tilde{\psi_i}|^2)\approx P - P|\tilde{\psi_i}|^2$. The violation of these conditions changes the dynamical behaviour by  shifting the boundaries of the states. Nevertheless, the overall transitions from symmetric bonding, symmetric antibonding,  asymmetric and oscillatory states of artificial dimers remain and can therefore be viewed as a generic feature of gain-dissipative oscillators existing in a large variety of physical systems.  
The couplings $J_{ij}$ in these systems depend on the pumping intensity, $P$, the blueshift, $g$, and the separation between the CCs. It can be approximated using analysis on the pairwise interactions between the condensates  \cite{ohadi2016nontrivial,lagoudakis2017polariton} as
\begin{equation}
\label{J_equation}
    J = J_{0}P^2\cos(\beta P\sqrt{g}),
\end{equation}
 where $\beta$ depends on the remaining system parameters and pumping geometry. Now we can see how the states change as the pumping intensities and pumping geometry vary as evident in the bifurcation diagram on Fig.~\ref{dyad_figure}, while the possible stationary states are shown schematically in Figs.~\ref{dyad_figure}(a)-(d). 
 
\begin{figure}[h!]
    \includegraphics[width=\columnwidth]{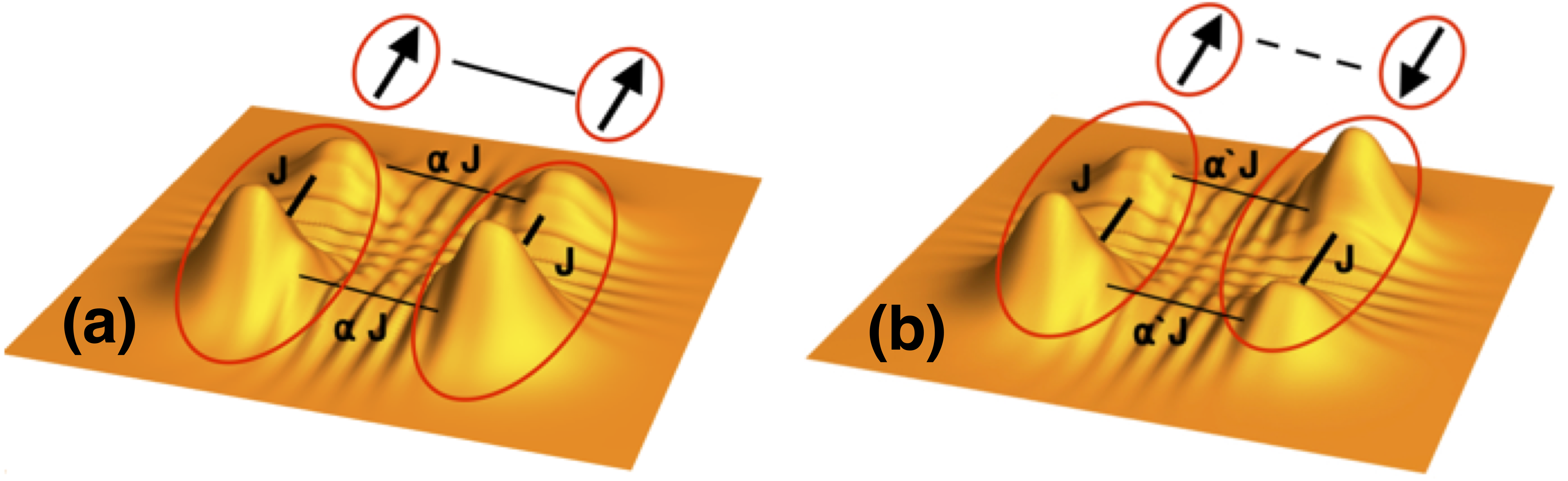}
    \caption{Interactions between two polariton dimers. Dimers are formed by bonds of strength $J>0$ while the couplings between dimers are much weaker: $\alpha J$ (a) and $\alpha' J$ (b), with $\alpha, \alpha'\ll 1.$ Increasing the weaker coupling strength above a critical threshold (here depicted by $\alpha \rightarrow \alpha'$) transitions the state transitions from $\uparrow \uparrow$ to $\uparrow \downarrow$, where the notion of direction is given by the density imbalance within a dimer. Relative strengths of the couplings are indicated (not to scale) by the thickness of the lines connecting the condensates. Here $J=0.5776$, $\beta = 2.5$, $J_{0} = 0.162$, $\alpha = 0.054$, $\alpha' = 0.056$, $b_{0} = 0.625$, $P = 2.52$, and $g=0.71$. For the $\uparrow \uparrow$ ($\uparrow \downarrow$) state higher densities are $ 2.6224$ ($2.5025$) and lower are $ 1.2731$ ($1.3904$).}
    \label{switching_state_diagram}
\end{figure}

 Asymmetric stationary states combine discrete and continuous degrees of freedom.  The continuous degree of freedom comes from the phase difference between "atoms". The discrete degree of freedom comes from occupation asymmetry that leads to non-trivial current  from one ``atom" to another.  In particular, we can associate Ising spins with the direction of this current in a polariton dimer. By changing the interactions between two dimers, we can create a pseudo-coupling between them so that the Ising spins align forming a type of ferromagnetic or antiferromagnetic coupling between the dimers. Such couplings are realised in Fig.~\ref{switching_state_diagram}. The ``atoms" from different ``molecules" are coupled less strongly (with the coupling strength $\alpha J$) than the ``atoms" within one molecule ($J$). By varying $0<\alpha \ll 1$ the orientation  of dimers change,  mimicking the behaviour of magnetic spins.



This combination of discrete observables associated with occupation asymmetry and continuous variables associated with the phase differences suggests applications in information processing. It has been argued that the intrinsic limitations of both discrete- and continuous-variable quantum information processing can be reduced by combining the two in a single platform \cite{andersen2015hybrid}. In our system, we  control  the values of the phases and the orientation of the spins through the adjustment of the coupling strength. It is possible to realise this on a large scale, in which each dimer forms the fundamental unit of an effective hybrid  information processing system. 

\begin{figure}[h!]
       \includegraphics[width=\columnwidth]{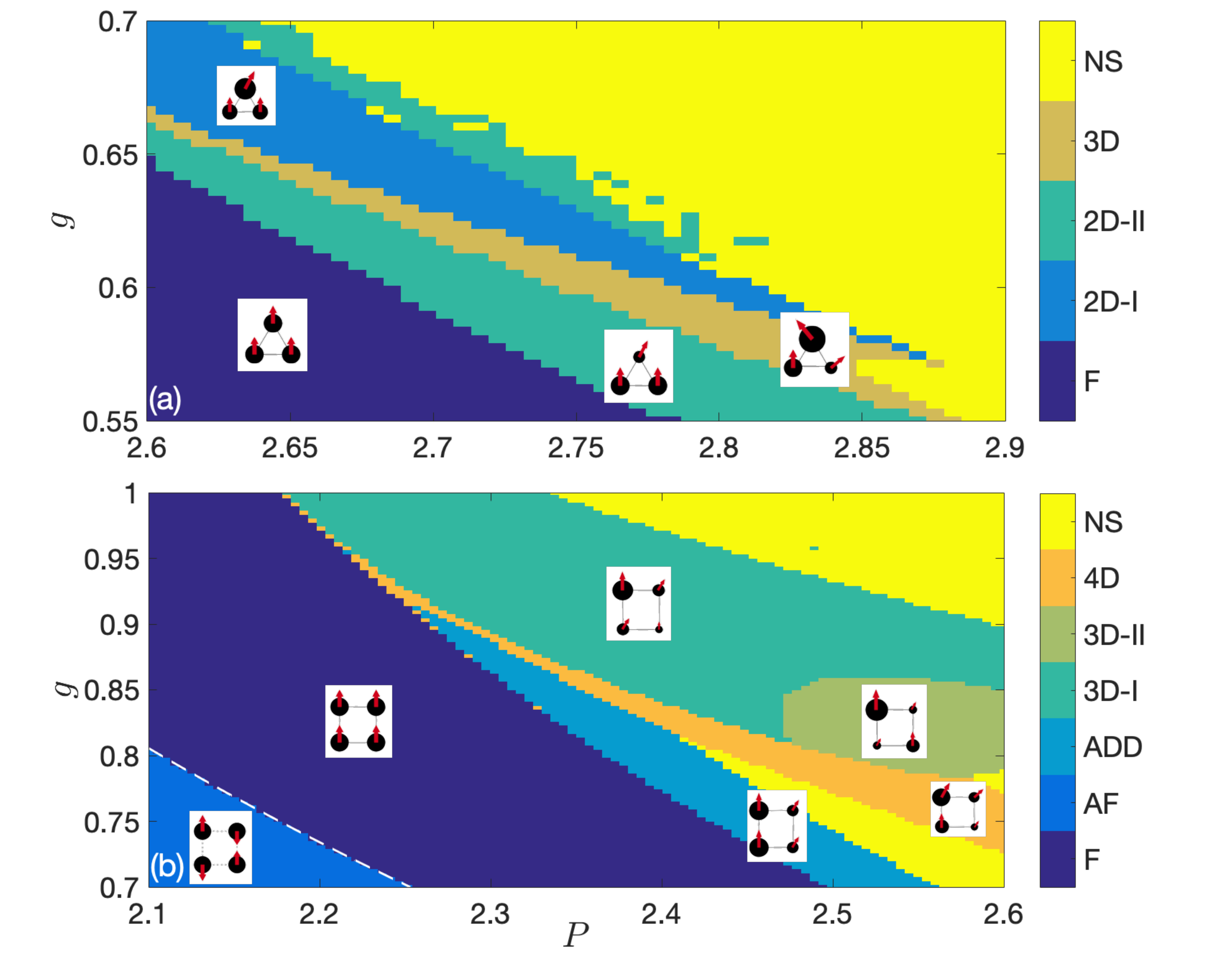}
     \caption{Bifurcation diagrams for polariton trimers (a)  and tetrads (b).  (a): symmetric bonding (ferromagnetic) (F), two density I (2D-I), two density II (2D-II),  three density (3D) states and non-stationary (NS) states; (b): symmetric bonding (ferromagnetic) (F), symmetric antibonding (antiferromagnetic) (AF), asymmetric double dimer (ADD), three density I (3D-I), three density II (3D-II),  four density (4D) states and  non-stationary (NS) states. The diagrams were calculated by numerical integration of  Eqs.~(\ref{wave_function_rate_equation}-\ref{reservoir_dynamics_equation}) with $J_{0} = 0.1,$ $\beta = 2.5$,  and $b_{0} = 0.5$ (a)  or  $b_{0} = 0.25$ (b).}
     \label{triad_tetrad_figure}
   \end{figure}
More complex molecular units are possible. Figure~\ref{triad_tetrad_figure} shows regions of parameter space in which all possible types of stationary states in polariton trimers (a) and tetrads (b)  are realised, as found by numerical integration of Eqs.~(\ref{wave_function_rate_equation}) and (\ref{reservoir_dynamics_equation}). In addition to the equal-density bonding and antibonding states, there are several distinct types of asymmetric stationary states comprised of  distinct densities. 
   

 
 The existence of four (six) independent steady states for a polariton tetrad suggests their potential use as the fundamental unit of a multistate logic system. Moreover, as Fig.~\ref{triad_tetrad_figure} illustrates, there exists a fixed $g$ that enables one to seamlessly transition between such states simply by changing $P$. Allowing the couplings to vary within a molecule further increases the possible number of discrete states in the system.

{\it Conclusions.} Nanotechnology has opened up the ability to create new classes of materials with designed properties. Controllable creation of intricate  artificial molecules that consist of internally coupled groupings of elementary oscillators  are testbeds for such new materials. In the Letter we demonstrated the range of configurations that can be assembled by coupling nonequilibrium polariton condensates. The physical properties of such configurations are significantly altered from those of the individual condensates due to the inter- and intra-condensate couplings. We show how the bonding and antibonding changes the spacing between the levels in dimers and brings about asymmetric states that  can be viewed as molecules with both discrete and continuous degrees of freedom. 
 By increasing the number of constituents we can introduce the basis of a polaritonic multi-valued logic system \cite{muthukrishnan2000multivalued}, such as those proposed for organic field-effect transistors \cite{kobashi2018multi}, optical \cite{ghosh2012trinary} and magnonic systems \cite{khitun2018magnonic}. 
 Oscillatory states of trimers and tetrads bring about frequency combs with controllable number of levels and spectral gaps, allowing one to manufacture the properties of artificial molecules by controlling the condensate geometry and coupling strengths. 
 Such artificial molecules can be readily manufactured using recent advances in freely expanding optically injected condensates in arbitrary graphs \cite{tosi2012geometrically,BerloffNatMat2017}, in photonic lattices with controlled loading of the condensate, in distinct orbital lattice modes of different symmetries, and at ambient conditions that include room temperature operation \cite{kena2010room,sun2017optical,dusel2019room}. 
 
\vspace{5mm}
 A.J. is grateful to Cambridge Australia Scholarships and the Cambridge Trust for wholly funding his PhD. K.P.K acknowledges support from Cambridge Trust and EPSRC. N.G.B. acknowledges support from Huawei.
 
\providecommand{\noopsort}[1]{}\providecommand{\singleletter}[1]{#1}%

\end{document}